\documentclass[11pt, a4paper]{article}
\pdfoutput=1
\usepackage[utf8]{inputenc}
\usepackage{graphicx}
\usepackage{geometry}
\usepackage[english]{babel}

\usepackage{tikz}
\usetikzlibrary{decorations.pathmorphing}
\usepackage{pgfplots}
\pgfplotsset{compat=1.15}
\usepgfplotslibrary{patchplots}
\usepackage{subcaption} 

\usepackage[numbers,sort&compress]{natbib}
\usepackage{epsfig}
\usepackage{amsmath}
\usepackage{amssymb}
\usepackage{amsfonts}
\usepackage{anysize}
\usepackage{latexsym}
\usepackage{xcolor}
\usepackage[pdftex,colorlinks=true,linkcolor=blue,citecolor=blue]{hyperref}
\hypersetup{linktocpage}
\usepackage{float}
\usepackage{dsfont}
\numberwithin{equation}{section}

\definecolor{ccqqqq}{rgb}{1,0.5,0}
\definecolor{uuuuuu}{rgb}{0.26666666666666666,0.26666666666666666,0.26666666666666666}
\definecolor{qqwwzz}{rgb}{0,0.3,0.9}

\newcommand{\beq}{\begin{equation}}
\newcommand{\eeq}{\end{equation}}
\newcommand{\bea}{\begin{eqnarray}}
\newcommand{\eea}{\end{eqnarray}}
\newcommand{\bit}{\begin{itemize}}
\newcommand{\eit}{\end{itemize}}

\def\s{\sigma}

\def\p{\partial}

\def\le{\left(}
\def\ri{\right)}

\date{}

\begin{document}

\begin{titlepage}

\begin{flushright}
%FTPI-MINN-08/06, UMN-TH-2638/08\\
%March 31\\

\end{flushright}
\bigskip
\begin{center}
{\LARGE  {\bf
Volume complexity for the non-supersymmetric Janus AdS$_5$ geometry
  \\[2mm] } }
\end{center}
\renewcommand{\thefootnote}{\fnsymbol{footnote}}
\bigskip
\begin{center}
 {\large \bf Stefano Baiguera},
  {\large \bf Sara Bonansea}, {\large \bf and } 
    {\large \bf Kristian Toccacelo}
\vskip 0.20cm
\end{center}
\vskip 0.20cm 
\begin{center}
{ \it \small{The Niels Bohr Institute, University of Copenhagen, \\
Blegdamsvej 17, DK-2100 Copenhagen \O, Denmark}}
\\ \vskip 0.20cm 
E-mails: stefano.baiguera@nbi.ku.dk, \\
sara.bonansea@nbi.ku.dk, btm438@alumni.ku.dk
\end{center}
\vspace{3mm}

\begin{abstract}
\begin{itemize}
We compute holographic complexity for the non-supersymmetric Janus deformation of AdS$_5$ according to the volume conjecture. The result is characterized by a power-law ultraviolet divergence. When a ball-shaped region located around the interface 
%separates the boundary into subregions, 
is considered,
a sub-leading logarithmic divergent term and a finite part appear in the corresponding subregion volume complexity. Using two different prescriptions to regularize the divergences, we find that the coefficient of the logarithmic term is universal.

\end{itemize}
\end{abstract}

\end{titlepage}

\tableofcontents

%%%%%%%%%%%%%%

\section{Introduction}

A major role in the development of theoretical physics in the last decades was played by the AdS/CFT correspondence, the most studied example being the duality between $\mathcal{N}=4$ super Yang-Mills (SYM) theory with gauge group $\mathrm{SU}(N)$ and type IIB string theory on $\mathrm{AdS}_5 \times S^5 $ \cite{Maldacena:1997re}.
While holography is a powerful tool in dealing with some particular physical systems, it is difficult to apply and test it in the strongly coupled phase of gravity.
An enormous improvement in the quantitative understanding of the correspondence has been possible thanks to the relationship between geometric objects in the bulk and information properties of quantum systems, starting from the duality between the entanglement entropy of a state on the boundary and the area of a codimension-two extremal surface \cite{Ryu:2006bv}. 
It was recently argued that the evolution of the Einstein-Rosen Bridge (ERB) cannot be captured by entropy, since it grows for a much longer timescale compared to the thermalization time.
For this reason, a new boundary quantity that supposedly encodes the information on the ERB has been introduced, that is, complexity.
% Thus, the new boundary quantity that is supposed to encodes the information on the ERB was proposed to be complexity. 
%For a quantum-mechanical system, it is defined as the minimal number of simple unitary operations which are needed to prepare a given state starting from a reference state.
Two different gravity duals for computational complexity have been conjectured: the complexity=volume (CV) \cite{Susskind:2014moa, Stanford:2014jda} and the complexity=action (CA) \cite{Brown:2015lvg, Brown:2015bva}.
In the CV conjecture, complexity is proportional to the volume of a maximal codimension-one submanifold hanging from the boundary
\beq
\mathcal{C}_V = \frac{\mathcal{V}}{G L} \, ,
\eeq
where $\mathcal{V}$ is the above-mentioned volume, $G$ the Newton constant and $L$ the AdS radius.
%In the CA conjecture, complexity is proportional to the gravitational action $I$ evaluated in the Wheeler-De Witt (WDW) patch, \emph{i.e.} the bulk domain of dependence of a Cauchy surface anchored at the boundary
In contrast, CA-duality relates the complexity on the boundary to the gravitational action $I$ evaluated on the Wheeler-De Witt (WDW) patch, i.e., the bulk domain of dependence of a Cauchy surface anchored at the boundary 
\beq
\mathcal{C}_A = \frac{I_{\rm WDW}}{\pi \hbar} \, .
\eeq
One of the original motivations to formulate the action proposal was the universality of the definition, since it does not require the introduction of an ad-hoc length scale. Early studies of the two conjectures for asymptotically AdS black holes have shown that the growth rate is the same at late times \cite{Brown:2015bva}.
However, it was later found that the proposals have different behaviors at intermediate times \cite{Lehner:2016vdi, Carmi:2017jqz}.
Many attempts to distinguish volume from action have been made by studying the following scenarios: the complexity of formation of a black hole from empty AdS space \cite{Chapman:2016hwi}, time-dependent spacetimes \cite{Moosa:2017yvt, Chapman:2018dem, Chapman:2018lsv}, cosmological models \cite{Barbon:2015ria}, backgrounds with non-relativistic traits \cite{Alishahiha:2018tep, Auzzi:2018zdu, Auzzi:2018pbc}, higher derivative gravity \cite{Alishahiha:2017hwg, Ghodrati:2017roz} and more.

%A parallel route from the boundary point of view was explored.

Parallel developments from the boundary theory perspective have also taken place. Computational complexity is defined in Quantum Mechanics (QM) as the minimal number of simple\footnote{The term simple usually refers to operators which are $k$--local, i.e., operators that act on at most $k\in\mathbb{N}$ qubits at the same time. The typical choice is to consider 2-local operators, since it is the simplest action that creates a non-vanishing entanglement.} unitary operators that must be used to transform a reference state into a target state.  
This definition is well-posed for discrete operators in quantum circuits, however, in view of the application to holography, it is necessary to generalize the definition to the continuum case. A possible route was proposed by Nielsen et al. in \cite{Nielsen1133}.
Adopting Nielsen's approach, many advances have been made both from the QM and QFT point of view, such as the investigation of curvature properties of the unitary group \cite{10.5555/2016985.2016986}, the relation between the spaces of unitaries and states in terms of Riemannian submersions \cite{Auzzi:2020idm}, the application to the SYK model \cite{Balasubramanian:2019wgd}, the properties of geodesics for integrable and chaotic systems \cite{Balasubramanian:2021mxo}, the implementation in free theories \cite{Jefferson:2017sdb, Chapman:2018hou, Khan:2018rzm}, the proposal of a first law of complexity \cite{Bernamonti:2019zyy} and the application to CFTs \cite{Caputa:2018kdj, Chagnet:2021uvi}.
A different avenue is that of path integral optimization \cite{Caputa:2017yrh, Boruch:2021hqs}.  

An intriguing problem is the investigation of the UV divergences that arise in the computation of complexity for both sides of the duality.
In the case of entanglement entropy, divergences occur as a consequence of the arbitrarily short correlations entangling the degrees of freedom in a subregion to the degrees of freedom in its complement.
A general classification reveals that the leading divergence scales with an area law.
Depending on the number of boundary dimensions, there is a universal term corresponding to the coefficient of the logarithmic divergence (even dimensions) or to the finite part (odd dimensions).
Such terms are universal because they do not depend on rescalings of the UV regulator.
This kind of general classification can also be done for the CV and CA conjectures, leading to the identification of the divergences in terms of curvature invariants integrated over a spatial slice \cite{Carmi:2016wjl, Reynolds:2016rvl}.

In this paper, we analyze the structure of divergences and their universality properties in the case of CV-duality applied to a defect geometry. A defect is generally defined as a modification of a system localized on a submanifold. Conformal defects provide a useful tool to probe the dynamics of some theories \cite{Cardy:1991tv, McAvity:1993ue, McAvity:1995zd}. Generally speaking, the insertion of a conformal defect in the vacuum of a theory results in the breaking of the full conformal group into a less constraining subgroup. 
%Several types of correlation functions which are forced to vanish in the absence of a defect due to the strong constraints imposed by conformal symmetry, become nontrivial when a defect is present. As a consequence, defect conformal field theories (dCFTs) require a broadened set of conformal data.
 Boundaries, defects, and interfaces have many applications both from a theoretical and phenomenological point of view. They constitute a simple path to bridge the gap between highly symmetric models studied in the context of the AdS/CFT duality and more physically realistic systems. For instance, condensed matter systems have impurities and, being finite in size, are also restricted by boundaries. Typical examples of defects are Wilson and 't Hooft operators in gauge theories \cite{Berenstein:1998ij,Kapustin:2005py} and D-branes in string theory. In high energy physics, defects can be engineered holographically \cite{Karch:2000ct,DeWolfe:2001pq}. For example, the study of the CFT data associated to the D3-D5 brane system described in \cite{Gaiotto:2008sa} was started in \cite{Nagasaki:2012re,deLeeuw:2015hxa} and a plethora of subsequent studies have been carried out \cite{Buhl-Mortensen:2015gfd,Buhl-Mortensen:2016pxs,Buhl-Mortensen:2016jqo,Buhl-Mortensen:2017ind,deLeeuw:2017dkd,Bonansea:2019rxh,Komatsu:2020sup,Bonansea:2020rkv}.
 
Furthermore, it is enticing to think of using defects as possible means to distinguish between the volume and the action proposals above-mentioned. This route was taken in \cite{Chapman:2018bqj}, where the CA and CV conjectures were inspected in the case of a bottom-up Randall-Sundrum type model \cite{Aharony:2003qf} of a thin $\rm{AdS}_2$ brane embedded in $\rm{AdS}_3$ spacetime. The remarkable output of this analysis is that, for the CV proposal, a new logarithmic divergence appears in the holographic complexity due to the presence of the defect, whereas the action computation is unaffected. 
%Strikingly, this was the first case in which the results of the holographic CV and CA proposals strongly disagreed. 
In \cite{Sato:2019kik, Braccia:2019xxi}, the holographic complexity was computed for a boundary CFT (BCFT). In particular, the analysis of \cite{Braccia:2019xxi} focuses on the gravitational dual of a boundary conformal field theory in two dimensions $(\rm{BCFT}_2)$. 
The authors found that the boundary distinguishes between CA and CV proposals in this case, too. In particular, a logarithmic divergence, whose coefficient depends on the boundary data, occurs in the CV computation. 
Instead, for CA this divergence drops out and the dependence on the boundary data occurs in the finite term. In \cite{Auzzi:2021nrj},  volume complexity was computed for a defect theory consisting of a Janus deformation of $\rm{AdS}_3$ spacetime.
It turns out that this setting admits precisely a logarithmic divergence whose coefficient is universal (i.e., independent of the regularization scheme and temperature of the configuration).

% Even in this deformed geometry a logarithmic divergence appears in the computation of the extremal volume. 
 Nevertheless, in \cite{Sato:2019kik} it was pointed out that the result for $\rm{AdS}_3/\rm{BCFT}_2$ cannot be extended to higher dimensions, which is not surprising given that gravity is known to have special features in three dimensions. In fact, in higher dimensional cases, the boundary complexity is non-vanishing both in CV and CA conjectures, and only contains a power-law
%  does not vanish even in the CA conjecture and the extremal volume computation does not show a logarithmic divergence, but a  
divergence which depends on the dimensionality of the spacetime.

In this paper, we focus on a non-supersymmetric (non-SUSY) Janus deformation of $\rm{AdS}_5$ spacetime to inspect the structure and the universality of UV divergences, in particular in comparison to the lower-dimensional case \cite{Auzzi:2021nrj}.
Volume complexity in the context of theories with defects was also recently studied in \cite{Flory:2017ftd, Bhattacharya:2021jrn}.

%if the logarithm divergence found in the computation of the CV proposal in \cite{AuzziJanus} for Janus-$\rm{AdS}_3$ is replaced by a power-law divergence.

The paper is organized as follows. In Section~\ref{Five-dimensional Janus AdS geometry} we describe the foliation of AdS$_{d+1}$ space into AdS$_d$ slices, which is useful to describe the defect geometry, and the properties of five-dimensional non-SUSY Janus spacetime. In Section~\ref{Volume for the Janus geometry}, we compute the complexity of formation for the Janus deformation using two different regularization schemes. In Section~\ref{sect-subregion_CV_AdS5}, we study the subregion complexity for a ball-shaped region centered on the defect using the same regularization procedures as for the total spacetime computation.
% with the two different regularization procedures used in the case of the total spacetime deformation. 
%We compare the results for the holographic complexity both in the subregion and the total cases to get insights into 
We comment on the universal structure of the UV divergences in the CV approach in Section \ref{sect-conclusions}. An appendix about the Weierstrass $\wp$--function completes the paper.

\section{Five-dimensional Janus AdS geometry}
\label{Five-dimensional Janus AdS geometry}
The non-SUSY Janus deformation of  $\mathrm{AdS}_5 \times S^5$  is a solution of type IIB supergravity with a non-trivial dilaton profile, which is regular and classically stable against all small and a certain class of large perturbations \cite{Bak:2003jk,Freedman:2003ax}. It can be thought of as a thick $\rm{AdS}_4$-sliced domain wall in $\mathrm{AdS}_5$.
 The CFT dual is given by the $\mathcal{N}=4$ Super Yang-Mills (SYM) theory on both sides  of a planar codimension-one interface, whose coupling constant varies discontinuously across the interface \cite{DHoker:2006vfr}  where the half-spaces are glued together. The two different values of the gauge coupling correspond to the two asymptotic values of the dilaton in the Janus deformation. 
 The $\mathrm{SO}(2,3)$ symmetry of the Janus geometry maps to the conformal group preserved by the three-dimensional interface on the CFT side. This symmetry is manifest at the classical level, 
 but was also shown to persist at the first non-trivial quantum level \cite{Clark:2004sb}. 
 The $\mathrm{SO}(6)$ symmetry of the Janus solution maps to an (accidental) internal symmetry\footnote{Since the Janus solution breaks all the supersymmetries, 
the global $\mathrm{SO}(6)$ symmetry is no longer an R-symmetry.}
on the CFT side. 
%In contrast to the defect conformal field theories, 
The interface
carries no degrees of freedom in addition to the ones inherited from 
%the bulk\footnote{Here, the term "bulk" refers to the $3+1$ dimensional ambient theory.} 
$\mathcal{N}=4$ SYM.
%, whence the name “interface”, as opposed to “defect”.  
 In this Section, we will introduce the basic information about the bulk geometries associated to theories with defects, referring in particular to the Janus interface solution.

\subsection{Geometries with defects}

An interface CFT with a codimension-one planar defect is invariant under the subgroup $\mathrm{SO}(d-1,2)$ of the original conformal group $ \mathrm{SO}(d,2)$. Its holographic dual is described by $\mathrm{AdS}_{d+1}$ space foliated into $\mathrm{AdS}_d$ slices with metric  \cite{Estes:2014hka, Gutperle:2016gfe}
\beq
ds^2 = L^2 \le  A^2 (y) ds^2_{\mathrm{AdS}_d} + \rho^2(y) dy^2 \ri \, ,
\label{eq:metric_Trivella_form}
\eeq
where $y$ is a non-compact coordinate. When $y \rightarrow \pm \infty$, we reach the asymptotic regions with the following behavior for the metric coefficients
\beq
A(y) \rightarrow \frac{L_{\pm}}{2} e^{\pm y \pm c_{\pm}} \, , \qquad
\rho(y) \rightarrow 1 \, .
\label{eq:asymptotic_Trivella_functions}
\eeq
In this context, $L_{\pm}$ and $c_{\pm}$ are constants which can in principle assume two different values at $y= \pm \infty.$
We parametrize the $\mathrm{AdS}_d$ slices using Poincaré coordinates 
\beq
ds^2_{\mathrm{AdS}_d} = \frac{1}{z^2} \le dz^2 - dt^2 + d \vec{x}^2_{d-2} \ri \, ,
\label{eq:metric_AdS_slicing}
\eeq
where $(t,z)$ are the time and radial coordinates on each slice and $\vec{x}$ identifies the other orthogonal spatial directions. 

Geometries of the kind described by Eq.~\eqref{eq:metric_Trivella_form} admit a Fefferman-Graham (FG) expansion close to the asymptotically $\mathrm{AdS}_{d+1}$ region of spacetime which brings the metric into the form
\beq
ds^2 = \frac{L^2}{\xi^2}  \left[ d \xi^2 + f_1 (\xi/\eta) \,  \le -dt^2 + d\vec{x}^2 \ri + f_2 (\xi/\eta) \, d\eta^2  \right]  \, .
\label{eq:form_metric_FGexpansion}
\eeq
Here, $\xi$ is a radial coordinate for the asymptotic $\mathrm{AdS}_{d+1}$ metric in Poincaré coordinates, $\eta$ is boundary direction orthogonal to the interface and $f_1, f_2$ are two functions encoding the change of coordinates.
% from the original metric \eqref{eq:metric_Trivella_form} with slicing \eqref{eq:metric_AdS_slicing} to this one.

Empty $\mathrm{AdS}_{d+1}$ space itself can be described in terms of an $\mathrm{AdS}_d$ slicing once we identify
$A(y)=\cosh y$ and $\rho(y)=1$ in Eq.~\eqref{eq:metric_Trivella_form}.
%We explain the procedure for empty $\mathrm{AdS}_{d+1}$ space, which can be written in terms of the slicing \eqref{eq:metric_Trivella_form} if we identify 
In this case, the FG expansion brings the metric into the exact Poincaré form
\beq
ds^2 = \frac{L^2}{\xi^2} \le d\xi^2 + d\eta^2 + d\vec{x}^2 - dt^2 \ri \, ,
\label{eq:metric_FG_Poincare_expansion}
\eeq
and the coordinate transformation reads
\beq
\eta= z \tanh y \, , \qquad \xi = \frac{z}{\cosh y} \, .
\label{eq:change_coordinates_FGform_emptyAdS}
\eeq
In the general case of a defect geometry, it may not be possible to find a closed form for the FG coordinate transformation, but an asymptotic expansion around the Poincaré solution can always be performed \cite{Estes:2014hka, Papadimitriou:2004rz}.
The FG expansion of the metric gives a natural prescription to regularize divergent quantities in the bulk geometry, as we will describe in Section \ref{sect-different_regularizations}.

\subsection{Non-supersymmetric Janus $\mathrm{AdS_5}$ geometry}

The non-SUSY five-dimensional Janus solution \cite{DHoker:2006vfr, Estes:2014hka} is a one-parameter deformation of AdS$_5$ described in terms of the metric
 \beq
\label{metric5}
ds^2 = L^2 \left[(\gamma)^{-1} h^2(w)  dw^2 + h(w) ds^2_{\mathrm{AdS}_4} \right] \, ,
\eeq
where $\gamma$ is the deformation parameter, with range 
$ 3/4 \leq \gamma \leq1  , $ and the four-dimensional AdS slice is written in Poincaré coordinates according to Eq.~\eqref{eq:metric_AdS_slicing}.
%\beq
%ds^2_{\mathrm{AdS}_4} = \frac{1}{z^2} \le dz^2 - dt^2  + d\vec{x}^2  \ri \, ,
%\eeq
%where $\vec{x}$ is a two-dimensional vector.
The warp factor $h(w)$ is defined as \cite{DHoker:2006vfr, Estes:2014hka}
\beq
\label{def_h_AdS_5}
h(w) = \gamma \le 1 + \frac{4 \gamma-3}{\wp(w)+1 -2\gamma} \ri 
=  \gamma \le 1 + \frac{4 \gamma-3}{\wp(w) - \wp(w_0)} \ri
 \, ,
\eeq
where $\wp(w)$ is the Weierstrass elliptic $\wp$--function\footnote{We refer the reader to Appendix \ref{app-weierstrass} for more details on the Weierstrass elliptic function.}.
The elliptic invariants $(g_2,g_3)$ of the Weierstrass $\wp-$function are
\beq
g_2 = 16 \gamma (1-\gamma) \, , \qquad
g_3 = 4(\gamma -1) \, ,
\eeq
and $w_0$ is defined as the positive solution of
\beq
\wp(w_0) = 2 \gamma -1 \, .
\eeq
The spatially varying dilaton of the non-SUSY Janus solution is
\beq
\phi(w)=\phi_0+\sqrt{6(1-\gamma)}\le w+\frac{4\gamma-3}{\wp^{'}(w)}\le \ln \frac{\sigma(w+w_1)}{\sigma(w-w_1)}-2\zeta(w_1)w \ri\ri\;,
\eeq
where $\s$ and $\zeta$ denote the Weierstrass functions defined in Eq.~\eqref{eq:zeta_sigma_functions} of Appendix \ref{app-weierstrass}, $\phi_0$
 is a real constant and $w_1$ is defined by the equation 
 \beq 
 \wp(w_1)=2(1-\gamma) \, .
 \eeq
When $\gamma=1$, the solution reduces to  $\mathrm{AdS}_5$ with constant dilaton $\phi=\phi_0$,  while $\gamma=3/4$ leads to a linear dilaton. 

The Janus solution is defined in the interval $- w_0 <w <w_0 $ since
% The values $\pm w_0$ are both solutions of $\wp (\pm w_0)=2\gamma-1$, 
% since $\wp(w)$ is an even function. 
 the function $h(w),$ introduced in Eq.~\eqref{def_h_AdS_5}, has simple poles at $w = \pm w_0$. 
 As $w \rightarrow \pm w_0$, the  Janus deformation asymptotes to AdS$_5$ with 
 constant dilaton $\phi_{\pm}=\phi(\pm w_0)$, where $\phi_{+}\neq \phi_{-}$  unless 
$\gamma = 1$. In other words, for generic $\gamma$ the  Janus solution has
 two asymptotically AdS$_5$ regions in which the dilaton takes two different values.

\begin{figure}[h]
\begin{center}
\begin{tikzpicture}[thick,scale=1.2]
\draw (-2,2) -- (0,0)
node[midway, left, inner sep=2mm] {$-\mu_0$};
\draw (2,2) -- (0,0)
node[midway, right, inner sep=2mm] {$\mu_0$};
\draw [dashed] (0,0) -- (0,-1.5);
\node (w) at (0,0.3) {$J$};

\end{tikzpicture}
\caption{Conformal diagram of the Janus $\mathrm{AdS}_5$ geometry with Poincaré coordinates on the $\mathrm{AdS}_4$ slices. 
The angular coordinate is ranged in the interval $[-\mu_0,\mu_0],$ where $\mu_0 > \pi/2,$ which gives the joint $J.$
It can be seen as an interface on the boundary.
In the diagram we suppress a factor of $\mathbb{R}^2$ at each point.}
\label{fig-conformal_diagram_Janus_AdS}
\end{center}
\end{figure}
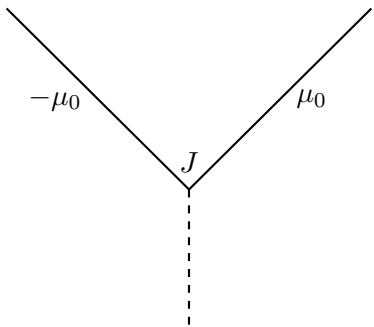

The conformal structure of the Janus $\mathrm{AdS}_5$ geometry is easily determined by means of the change of variables
\beq
d \mu = \sqrt{\frac{ h (w)}{\gamma}} \, dw  \, ,
\eeq 
which brings the metric into the form
\beq
ds^2 = L^2 \, h(\mu) \le  d\mu^2 + ds^2_{\mathrm{AdS}_4} \ri \, .
\eeq
Up to a conformal factor, the boundary metric is four-dimensional Minkowski spacetime.
The conformal diagram corresponding to this geometry is shown in Fig.~\ref{fig-conformal_diagram_Janus_AdS}.

%%%%%%%%%%%%%%%%%%%%%%%

\section{Volume for the Janus $ \mathrm{AdS}_5 $ geometry}
\label{Volume for the Janus geometry}
In this Section, we study the holographic complexity according to the CV conjecture in the non-SUSY Janus deformed $\mathrm{AdS}_5$ space. 
In the three-dimensional case, the extremal volume of the entire space for both the Janus $\mathrm{AdS_3}$ \cite{Auzzi:2021nrj} and the Randall-Sundrum \cite{Chapman:2018bqj} defect models exhibit a logarithmic divergence. 
In \cite{Sato:2019kik} it is pointed out that, for a boundary CFT (BCFT), the behavior of the divergent terms for the CV conjecture depends on the dimensionality of the space. Since a BCFT can be related to a CFT with a codimension-one defect via the unfolding trick \cite{Bachas:2001vj}, we expect an analogous behavior also for the Janus interface. 

\subsection{UV regularizations of the extremal volume}
\label{sect-different_regularizations}

A common feature of geometrical objects defined via the AdS/CFT correspondence is the existence of UV divergences that need to be regularized to describe physically meaningful quantities.
This happens, e.g., for the computation of the entanglement entropy, the extremal volume identified by the CV conjecture, or the on-shell gravitational action which plays a role in the computation of free energy or the CA conjecture.
There exist essentially three regularization techniques that allow to extract the same physical information \cite{Gutperle:2016gfe}.
Here, we review the main aspects of these methods.

The extremal volumes that we will compute in this Section are of the form
\beq
\mathcal{V} = \int dz \int dy \int d\vec{x} \, \sqrt{h} \, ,
\label{eq:general_prescription_volume}
\eeq
where $h$ is the determinant of the induced metric on the codimension-one Cauchy spatial slice.
The integrations along the $\vec{x}$ directions are always trivial, while the relevant information on the interface are encoded by the $(y,z)$ directions.
\\
\vspace{1mm} \\
\textbf{Fefferman-Graham regularization.} 
A natural way to regularize UV divergences is provided by the FG metric \eqref{eq:form_metric_FGexpansion}. 
It consists in cutting the geometrical object of interest with the hypersurface located at $\xi= \delta,$ and all the results are expressed in terms of a series expansion around $\delta =0.$
The problem of this procedure is that in the region where $\xi/\eta \ll 1,$  the FG expansion breaks down and the coordinates $(\xi, \eta)$ are ill-defined \cite{Papadimitriou:2004rz}.

\begin{figure}[h]
\begin{center}
\begin{tikzpicture}[thick,scale=1.3]
\fill[red!40!yellow] (0,0) -- (-0.75,3) -- (0.75,3) -- (0,0);
\fill[black!20!white] (0,0) -- (-5,0) -- (-5,3) -- (-0.75,3) -- (0,0);
\fill[black!20!white] (0,0) -- (5,0) -- (5,3) -- (0.75,3) -- (0,0);
\draw (-5,0) -- (5,0)
node[midway, below, inner sep=2mm] {\textit{Defect}};
\draw (-5,1) -- (-0.26,1)
node[midway, below, inner sep=1mm] {$\xi=\delta$}
node[midway, above, inner sep=9mm] {\textit{Left FG Patch}};
\draw (0.26,1) -- (5,1) 
node[midway, below, inner sep=1mm] {$\xi=\delta$}
node[midway, above, inner sep=9mm] {\textit{Right FG Patch}};
\draw (0.26,1) arc (45:135:0.37)
node[midway, above, inner sep=1mm] {\textit{$\Gamma$}};

\draw (0,0) -- (-0.75,3);
\draw (0,0) -- (0.75,3);
%\fill[blue!60!white,draw=black] (-1.5,0) -- (0,1.5) -- (1.5,0)--(0,-1.5);
\draw [black,fill] (0,0) circle [radius=0.07]; % Draws a circle
%\node (I)    at ( 0,0)   {\textit{Shadow region}};
\end{tikzpicture}
\caption{Interpolation between two FG patches with a continuous curve $\Gamma$. The black line, separating the grey region from the orange one (where the FG coordinates are ill-defined), corresponds to the value $y_0$ for the $y$-coordinate.
}
\label{fig-interpolation_FG_patches}
\end{center}
\end{figure}
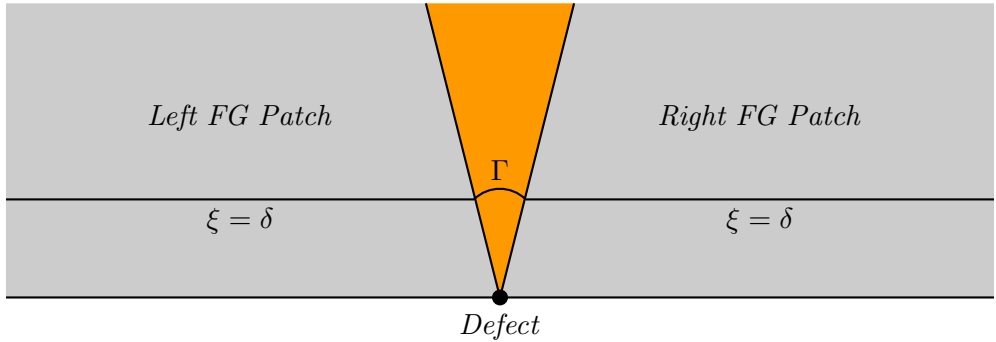

Indeed, the defect geometry contains two FG patches located away from the interface on its left and right sides.
In the middle region, there is not a well-defined change of variables that selects a natural UV cutoff.
This problem can be solved by interpolating the cutoff surface located at $\xi=\delta$
  in the left and right FG patches with an arbitrary curve in the middle region, with the constraint that it has to be smooth at  $y=y_0,$ which corresponds to the breaking down of the FG expansion \cite{Estes:2014hka, Chapman:2018bqj}.
This setting is depicted in Fig.~\ref{fig-interpolation_FG_patches}.
\vspace{1mm}  \\

\noindent
\textbf{Single cutoff regularization.} 
This method makes use of the FG expansion of the metric to identify a UV cutoff in the asymptotic region. 
Instead of performing an arbitrary interpolation in the middle region, it induces a minimal value for the $z$-coordinate. 

We explain the details of the procedure starting from the empty AdS case, whose FG expansion is achieved by means of the coordinate transformation \eqref{eq:change_coordinates_FGform_emptyAdS}.
If we locate the UV cutoff at the surface $\xi=\delta,$ we get the condition
\beq
\delta = \frac{z}{\cosh y}  \, ,
\label{eq:one_cutoff_FG_empty_AdS}
\eeq
which selects a maximal value of $y=y^*(z)$  in Eq.~\eqref{eq:general_prescription_volume}.
On the other hand, reversing this identity gives a constraint on the minimal value of the $z,$ determined by
\beq
z_{\rm min} = \delta \,  \underset{y \in \mathbb{R}}{\mathrm{min}} \le  \cosh y \ri = \delta  \, .
\eeq
In the presence of an interface, the procedure is formally the same, but the conditions to impose become
\beq
 \delta=\frac{z}{A(y)}  \, , \qquad
z_{\rm min} = \delta \, \underset{y \in \mathbb{R}}{\mathrm{min}} \, [A(y)] \, ,
\label{eq:single_cutoff_prescription}
\eeq
where $A(y)$ was introduced in Eq.~\eqref{eq:metric_Trivella_form}.
This choice degenerates to the empty AdS case once the deformation parameter of the Janus geometry is turned off.
At the end of the procedure, we will perform a Laurent expansion of the result in powers of $\delta.$
\\
\vspace{1mm} \\
\textbf{Double cutoff regularization.}
Another way to regularize an integration along the two directions $(y,z)$ is to introduce two separate UV cutoff.
A natural choice is to impose $z= \delta$ on the $\mathrm{AdS}_d$ slicing  instead of making use of the FG expansion.
Since the metric factor $A(y)$ is singular at infinity even after this regularization, a maximum value of $y$ is determined by  requiring 
\beq
A(y) = \frac{1}{\varepsilon} \, ,
\label{eq:double_cutoff_prescription} 
\eeq 
which is a natural counterpart of Eq.~\eqref{eq:single_cutoff_prescription}.
Notice that while the $\delta$ cutoff has physical relevance as it regularizes the intrinsic contribution from the defect to the volume, the $\varepsilon$ cutoff is a mathematical artifact introduced at intermediate steps. This parameter is sent to zero at the end of the computation, and the physical quantities will not depend on it.

\subsection{Extremal volume: single cutoff procedure}
\label{subsec_extremal_volume_AdS_5}

We evaluate the extremal volume using the metric in Eq.~\eqref{metric5}.  Since the integral that defines the volume diverges near the asymptotic boundary, i.e., when $w \rightarrow \pm w_0$ and $z \rightarrow 0$, we have to regularize it introducing suitable cutoffs. 
We adopt the single cutoff regularization procedure described in Section \ref{sect-different_regularizations}, which relies on the intertwining between the UV cutoff along the $w$ and $z$ directions.

\subsubsection*{Determination of the geometric data}
 
First of all, we need to re-write the metric of Janus $\mathrm{AdS}_5$ space into the general form given in Eq.~\eqref{eq:metric_Trivella_form}, where the coordinate $y$ is non-compact and the prefactor of the $dy^2$ terms is the unity.
This can be easily achieved by performing the following change of coordinates
\beq
dy= \gamma^{-1/2} h (w) dw \quad
\Rightarrow \quad
y = \gamma^{-1/2} \int_0^{w} dw' \, h(w') \, ,
\eeq
which brings the metric \eqref{metric5} into the form
\beq
ds^2 = L^2 \le dy^2 + h(y) ds^2_{\mathrm{AdS}_4} \ri \, ,
\eeq
where we identify 
\beq
A(y)=\sqrt{h(y)}\;, \qquad \rho(y) = 1\;.
\eeq
According to Eq.~\eqref{eq:single_cutoff_prescription}, the cutoff surface at $\xi=\delta$ gives the constraint
\beq
\delta = \frac{z}{\sqrt{h(y)}} \, ,
\label{eq:single_cutoff_condition_JanusAdS5}
\eeq
which consequently induces the value of $z_{\rm{min}}$ as
\beq
\label{z_min_AdS_5}
z_{\rm min} = \delta \,  \underset{y \in \mathbb{R}}{\mathrm{min}} \le \sqrt{h(y)} \ri  = \delta \, \sqrt{\gamma} \;,
\eeq
where we used that $h$ takes minimum value at $y=0,$ and $h(0)=\gamma .$  
The previous prescription  can be equivalently employed using the compact coordinate $w$ and it defines a cutoff $w_{\pm}$ such that
\beq
h(w_{\pm}) = \frac{z^2}{\delta^2}  \quad \Rightarrow \quad
w_{\pm} = h^{-1} \le \frac{z^2}{\delta^2} \ri \, .
\label{eq:condition_wpm}
\eeq
It regularizes the divergencies stemming from the poles of the function $h(w)$ located at $w=\pm w_0.$
The cutoff $w_{\pm}$ can be expanded in a power series of $\delta/z,$ as shown in \cite{Estes:2014hka}
\beq
\label{expansion_w_extrema}
w_{\pm}\left(\frac{\delta}{z} \right) =\pm w_0 \mp \sum_{k=1}^{\infty} b_k \, \frac{\delta^{2k}}{z^{2k}} \, .
\eeq
All the coefficients of the series can be recursively determined order by order by imposing the condition \eqref{eq:condition_wpm}.
The previous expression only contains even powers of $\delta/z$ because $h(w)$ is an even function, and the first coefficients of the series are given by
\beq
b_1 =  \frac{\sqrt{\gamma}}{2} \, , \qquad
b_2 = \frac{\sqrt{\gamma}}{8} \, , \qquad
b_3 =  \frac{\sqrt{\gamma}}{16} \, , \qquad
b_4 = \frac{5 \sqrt{\gamma}}{128} \, , \qquad \dots
\label{eq:first_coefficients_series_w_extrema}
\eeq
Whereas the location of the cutoff is fixed via a Taylor expansion, the identity \eqref{eq:condition_wpm} is exact and formally resums all the coefficients of the above series.

\subsubsection*{Computation of the volume}

It is not restrictive to study the CV conjecture in the deformed $\mathrm{AdS}_5$ background using a time slice at zero boundary time\footnote{It can be shown that the extremal slice at constant time is always a solution of the equations of motion at all times. By translational invariance along the time direction, we choose for convenience to study the case with vanishing boundary time.}.
The extremal volume is computed by the integral 
\beq
\mathcal{V}= \frac{2 L^4 V_2}{\sqrt{\gamma}}\int^{z_{\rm IR}}_{ \delta \sqrt{\gamma}} \frac{dz}{z^3}\int_{0}^{w_{+}\left( \frac{\delta}{z}\right) }h(w)^{\frac{5}{2}} dw\;,
\label{eq:volume_one_cutoff_integration_part1}
\eeq
where we introduced a factor of 2 because $\wp(w)$ is even, and we denoted with  $V_2$ the two-dimensional infinite volume along the orthogonal spatial directions.
Since the integral along the $z$ direction is in principle divergent at infinity, we regularize it introducing a cutoff $z_{\rm IR}.$ 

\begin{figure}
\begin{center}
\begin{tikzpicture}[thick,scale=1]
\draw [-stealth] (0,0) -- (6,0) node[right] {$\zeta$};
\draw [-stealth] (0,0) -- (0,5) node[above] {$\tau$};
\fill [red!80!yellow] (1,1) -- (4.5,4.5) -- (4.5,1) -- (1,1);
\draw [dashed] (1,0) -- (1,5);
\draw [dashed] (0,1) -- (5,1);
\node at (1,-0.3) {$1$};
\node at (-0.3,1) {$1$};
\node at (4.5,-0.3) {$\zeta_\text{IR}$};
\draw [dashed] (4.5,5) -- (4.5,0);
\draw (0,0) -- (4.5,4.5);
\draw [dashed] (4.5,4.5) -- (5,5);
\draw [dashed] (5,4.5) -- (0,4.5);
\node at (-0.3,4.5) {$\zeta_\text{IR}$};
\node at (2.5,2.8) [rotate=45] {$\tau=\zeta$};
\end{tikzpicture}
\caption{The domain of integration is depicted in red. We can swap the order of the integrals by adjusting the extremes of integration.}
\label{integration_domain}
\end{center}
\end{figure}
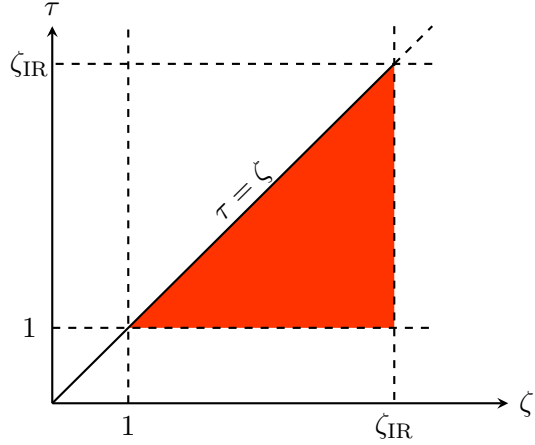

To evaluate the extremal volume, we begin with the change of variables $\tau = \frac{h(w)}{\gamma}.$  
%The transformation of the measure is evaluated using the following identities:
%\begin{itemize}
%\item The inverse of the change of variables.
%%\beq
%%\wp = \frac{2+ \tau - 2 \gamma ( \tau +1)}{1 - \tau} \, .
%%\eeq
%\item The differential of Eq.~\eqref{def_h_AdS_5}.
%%\beq
%%h'(w) = - \frac{\gamma(4 \gamma-3)}{\le \wp + 1 - 2 \gamma  \ri^2} \, \wp'(w) \, .
%%\eeq
%\item The property \eqref{eq:differential_eq_WeierstrassP} of the Weierstrass $\wp-$function, which gets rid of all the $\wp'$ terms.
%\end{itemize}
The extremes of integration are fixed by the conditions $h(0)= \gamma$ and $h(w_{\pm})= z^2 / \delta^2 .$ Using the properties of the Weierstrass $\wp-$function summarized in Appendix~\ref{app-weierstrass},
we get
\beq
\int_0^{w_+(\varepsilon)} dw \, h^{5/2} (w)  =  \frac{\gamma^{5/2}}{2}
\int_{1}^{\frac{z^2}{\gamma \delta^2}} d\tau \, 
\sqrt{\frac{\tau^5}{\gamma (\tau^4-1) - (\tau^3-1)}} \, .
\eeq
At this point, we perform a further change of variables 
$
\zeta = z^2/(\gamma \delta^2) ,$
which brings the volume \eqref{eq:volume_one_cutoff_integration_part1} into the form
\beq
\mathcal{V} = \frac{\gamma L^4 V_2}{2 \delta^2} \int_1^{\zeta_{\rm IR}} \frac{d\zeta}{\zeta^2} 
\int_1^{\zeta} d\tau \, \tau^{5/2} f(\tau) \, ,
\eeq
where we define
\beq
\label{def_func_f}
f(\tau) \equiv \frac{1}{\sqrt{\gamma(\tau^4-1)-(\tau^3-1)}} \, , \qquad
\zeta_{\rm IR} \equiv \frac{z_{\rm IR}^2}{\gamma \delta^2} \, .
\eeq
We can swap the order of integration as shown in Fig.~\ref{integration_domain} in the following way
\beq
\label{swapping}
\int_1^{\zeta_{\rm IR}} d \zeta \int_1^{\zeta} d\tau \, F(\tau,\zeta) \rightarrow
\int_1^{\zeta_{\rm IR}} d\tau \int_{\tau}^{\zeta_{IR}} d \zeta \, F(\tau,\zeta) \, ,
\eeq
for any given integrand function $F(\tau,\zeta).$ 
The evaluation of the $\zeta$ integration is trivial and gives
\beq
\mathcal{V} = \frac{\gamma L^4 V_2}{2 \delta^2} \int_1^{\zeta_{\rm IR}} d\tau \, \tau^{5/2} f(\tau) \le \frac{1}{\tau} - \frac{1}{\zeta_{\rm IR}} \ri \, .
\eeq
The remaining integral is divergent in the limit $\delta \rightarrow 0,$ i.e., $\zeta_{\rm IR} \rightarrow \infty.$
However, the function $f(\tau)$ can be series-expanded around infinity, where it is analytic.
We can add and subtract the lowest orders of the $f(\tau)$ expansion 
\beq
\begin{aligned}
\tau^{5/2} f(\tau) \le \frac{1}{\tau} - \frac{1}{\zeta_{\rm IR}} \ri & = 
\tau^{3/2} \left[\le  f(\tau) - \frac{1}{\sqrt{\gamma} \tau^2} - \frac{1}{2 \gamma^{3/2} \tau^3} \ri  + \frac{1}{\sqrt{\gamma} \tau^2} + \frac{1}{2 \gamma^{3/2} \tau^3}  \right] \\
& - \frac{\tau^{5/2}}{\zeta_{\rm IR}}  \left[ \le  f(\tau) - \frac{1}{\sqrt{\gamma} \tau^2} - \frac{1}{2 \gamma^{3/2} \tau^3}  \ri + \frac{1}{\sqrt{\gamma} \tau^2} + \frac{1}{2 \gamma^{3/2} \tau^3}   \right] \, ,
\end{aligned}
\eeq
in such a way that the terms in the round parenthesis define a  renormalized finite integral where the limit $\delta \rightarrow0$ can be performed directly.
Hence, we define 
\beq
A(\gamma) \equiv \int_1^{\infty} d\tau \, \tau^{3/2} \le  f(\tau) - \frac{1}{\sqrt{\gamma} \tau^2} - \frac{1}{2 \gamma^{3/2} \tau^3} \ri \, ,
\label{eq:definition_A_gamma}
\eeq
\beq
B(\gamma) \equiv \int_1^{\infty} d\tau \, \tau^{5/2}
\le  f(\tau) - \frac{1}{\sqrt{\gamma} \tau^2} - \frac{1}{2 \gamma^{3/2} \tau^3} \ri \, .
\label{eq:definition_B_gamma}
\eeq
%These functions are finite and well-defined, and in particular it is easy to expand them around $\gamma=1,$ since all the coefficients of the Taylor series can be computed analitically order by order.
%For small values of $(\gamma-1),$ the following expansion is trustable:
%\beq
%\label{A_function}
%A(\gamma) = 1 + \le \frac{1}{2} - \frac{15 \pi}{16} \ri (\gamma-1) + \frac{1}{2} \le \frac{5169 \pi}{1024} - \frac{9}{4} \ri (\gamma-1)^2 + \mathcal{O}((\gamma-1)^3)
%\eeq
%\beq
%\label{B_function}
%B(\gamma)= \frac{5}{3} - \le \frac{11}{6} + \frac{3 \pi}{4} \ri (\gamma-1) + \frac{1}{2} \le \frac{2697 \pi}{512} + \frac{17}{4} \ri (\gamma-1)^2 
%+ \mathcal{O}((\gamma-1)^3)
%\eeq
These functions can be numerically evaluated and are well-defined everywhere except near $\gamma = 3/4,$ which is a singular limit since  $w_0 \rightarrow \infty.$

%\begin{figure}[ht]
%    \centering
%    \def\svgwidth{\columnwidth}
%    \scalebox{0.65}{\input{BandA.pdf_tex}}
%  	\caption{Numerical plot of the functions $A(\gamma)$ and $B(\gamma)$ defined in Eqs.~\eqref{eq:definition_A_gamma} and \eqref{eq:definition_B_gamma}. The limit $\gamma \rightarrow 3/4$ is singular, while the limit $\gamma \rightarrow 1$ can be performed analitically from the Taylor expansion around that point, giving $A(1)=1$ and $B(1)=5/3.$}
%\label{fig-A_B_functions}
%\end{figure} 

In the following, we will keep implicit the functions $A(\gamma)$ and $B(\gamma).$ 
The divergences in the extremal volume are encoded by the following remaining terms
\beq
\int_1^{\zeta_{\rm IR}} d\tau \, \le \frac{1}{\sqrt{\gamma \tau}} + \frac{1}{2 \gamma^{3/2} \tau^{3/2}} \ri = 
\frac{1}{\gamma^{3/2}} \le 2 \gamma \sqrt{\zeta_{\rm IR}} - \frac{1}{\sqrt{\zeta_{\rm IR}}} + 1 - 2 \gamma \ri \, ,
\label{eq:integral1_onecutoff}
\eeq
\beq
\int_1^{\zeta_{\rm IR}} d\tau \, \le \sqrt{\frac{\tau}{\gamma}} + \frac{1}{2 \gamma^{3/2} \sqrt{\tau}} \ri = 
\frac{1}{\gamma^{3/2}} \le  \frac{2 \gamma}{3} \zeta_{\rm IR}^{3/2} + \sqrt{\zeta_{\rm IR}} - \frac{2}{3} \gamma -1 \ri \, .
\label{eq:integral2_onecutoff}
\eeq
Summing all the contributions with the appropriate pre-factors, we obtain the expression for the extremal volume
\beq
\mathcal{V} = L^4 V_2 \left[  \frac{2}{3} \frac{z_{\rm IR}}{\delta^3} + \le \frac{\gamma A(\gamma)}{2} -  \sqrt{\gamma}+ \frac{1}{2\sqrt{\gamma}} \ri \frac{1}{\delta^2} 
- \frac{1}{z_{\rm IR} \delta} - \frac{1}{2 z_{\rm IR}^2} \le \gamma^2 B(\gamma) -  \sqrt{\gamma} - \frac{2}{3} \gamma^{3/2} \ri  \right]
+ \mathcal{O}(\delta) \, .
\label{eq:volume_onecutoff_JanusAdS5}
\eeq
%Notice that the finite part of this expression is inversely proportional to $z_{\rm IR},$ which is the IR cutoff.
%Since the limit $z_{\rm IR} \rightarrow \infty$ can be performed smoothly, we directly apply it to achieve the result
%\beq
%\mathcal{V} = V_2 \left[  \frac{2}{3} \frac{z_{\rm IR}}{\delta^3} + \le \gamma A(\gamma) + \frac{1}{\sqrt{\gamma}} - 2 \sqrt{\gamma} \ri \frac{1}{\delta^2} 
%- \frac{1}{z_{\rm IR} \delta}  \right]
%+ \mathcal{O}(\delta) \, .
%\eeq
%We notice that in this way we got rid of the function $B(\gamma).$

\subsubsection*{Subtraction of vacuum AdS$_5$}
To identify the contribution of the defect to the extremal volume, we need to subtract the result that stems from vacuum $\mathrm{AdS}_5$. First of all, we write the $\mathrm{AdS}_5$ metric in $\mathrm{AdS}_4$ sliced form by setting $\gamma \rightarrow 1 $ in Eq.~\eqref{metric5}, recognizing that in this limit
\beq
h(w) = \frac{1}{1-w^2}, \qquad w_0=1\;.
\eeq   
Thus, the $\mathrm{AdS}_5$ metric can be re-written as
\beq
\label{AdS_5_metric_vacuum_CV}
ds^2_{\mathrm{AdS}_5}=\frac{L^2 dw^2}{(1-w^2)^2}+\frac{L^2}{1-w^2}ds^2_{\mathrm{AdS}_4}\;.
\eeq
Notice that the same result can be achieved by performing the change of coordinates
\beq
w=\tanh y
\eeq
in the standard unit-radius $\mathrm{AdS}_5$ metric written in Poincaré slicing
\beq
ds^2_{\mathrm{AdS}_5}=dy^2+\cosh^2\!y \,ds^2_{\mathrm{AdS}_4}\;.
\eeq
%% We have already encountered this precise transformation in the $\mathrm{AdS}_3$ case relating Eq.~\eqref{eq:AdS_2_slicing} to Eq.~\eqref{eq:AdS_2_slicing_2}. 
 The FG change of coordinates \eqref{eq:change_coordinates_FGform_emptyAdS} for the metric \eqref{AdS_5_metric_vacuum_CV} becomes
 \beq
 \label{FG_vacuum_AdS_5}
 \xi=z\sqrt{1-w^2}, \qquad \eta=zw\;,
 \eeq
and, by setting $\xi=\delta,$  they induce a cutoff $w_{*}$ given by
 \beq
 w_{*}=\sqrt{1-\frac{\delta^2}{z^2}}\;.
 \eeq
 The value of $z_{\rm min}$ is defined as
 \beq
 z_{\rm min}= \rm min \left(\frac{\delta}{\sqrt{1-w^2}} \right) =\delta\;.
 \eeq
 The extremal volume at $t=0$ for the $\mathrm{AdS}_5$ vacuum solution reads
 \beq
 \label{volume_vacuum_AdS_5}
 \mathcal{V}_{\mathrm{AdS}_5}=2 L^4 V_2\int_{\delta}^{z_{\rm IR}}\frac{dz}{z^3}\int_{0}^{\sqrt{1-\frac{\delta^2}{z^2}}}\frac{dw}{(1-w^2)^{5/2}} =
 2 L^4 V_2\left[\frac{1}{3}\frac{z_{\rm IR}}{\delta^3}-\frac{1}{2}\frac{1}{z_{\rm IR}\delta} \right] +\mathcal{O}(\delta)\;.
 \eeq

\subsubsection*{Total result} 
 
Once we subtract the volume for the empty $\mathrm{AdS}_5$ space to Eq.~\eqref{eq:volume_onecutoff_JanusAdS5}, we obtain the complexity of formation of the Janus $\mathrm{AdS}_5$ solution as
 \beq
 \Delta \mathcal{\mathcal{C}}_{\mathrm{AdS}_5} (\gamma)= \frac{V_2 L^3}{G \delta^2} \le \frac{\gamma A(\gamma)}{2} - \sqrt{\gamma}+ \frac{1}{2\sqrt{\gamma}} \ri 
- \frac{V_2 L^3}{2 G z_{\rm IR}^2} \le \gamma^2 B(\gamma)-  \sqrt{\gamma} - \frac{2}{3} \gamma^{3/2} \ri  
  +\mathcal{O}(\delta)\;. 
\label{eq:total_volume_JanusAdS5}
 \eeq 
This result shows that the contribution to the complexity intrinsic to the interface amounts to a power-law divergence $\delta^{-2}$ and to a finite part, which vanishes in the limit $z_{\rm IR} \rightarrow \infty.$
We will comment on this structure with further details after having studied the complexity of formation with the double cutoff regularization scheme.

\subsection{Extremal volume: double cutoff procedure}
\label{sect-Extramal_volume_2cutoff}

We employ the double cutoff prescription to regularize the UV divergences of the extremal volume.
Since we introduce two different regulators for the $z$ and $w$ directions, the integrations are not nested.
We select the UV cutoff along the $z$ direction to be $z=\delta,$  whereas to determine a cutoff in the $w$ variable we use
\beq
h(w_{\pm}) = \frac{1}{\varepsilon^2} \, ,
\eeq
according to Eq.~\eqref{eq:double_cutoff_prescription}.
%In this way the Taylor expansion \eqref{expansion_w_extrema} defining the location of the cutoff becomes
%\beq
%\label{expansion_w_extrema_2cutoff}
%w_{\pm}\left(\varepsilon \right) =\pm w_0 \mp \sum_{k=1}^{\infty} b_k \, \varepsilon^{2k} \, ,
%\eeq
%with the same coefficients determined in Eq.~\eqref{eq:first_coefficients_series_w_extrema}.

\subsubsection*{Computation of the volume}

The extremal volume at vanishing boundary time is determined by
\beq
\mathcal{V}= \frac{2 L^4 V_2}{\sqrt{\gamma}}\int^{z_{\rm IR}}_{ \delta} \frac{dz}{z^3}\int_{0}^{w_{+}\left( \varepsilon \right) }h(w)^{\frac{5}{2}} dw\;.
\eeq
Changing variables into $\tau=h(w)/\gamma$ and performing the integration over $z,$ we obtain
\beq
\mathcal{V} =  \frac{\gamma^2 L^4 V_2}{2 \delta^2} 
\le \frac{1}{2 \delta^2} - \frac{1}{2 z_{\rm IR}^2} \ri
\int_1^{\frac{1}{\gamma \varepsilon^2}} d\tau \, \tau^{5/2} f(\tau) \, .
\label{eq:integral_volume_2cutoff}
\eeq
The last integration in $\tau$ is carried out using the same method described in Section \ref{subsec_extremal_volume_AdS_5}.
Thus, the extremal volume in the double cutoff regularization scheme reads
%again by renormalizing the integrand at infinity, and evaluating the divergent parts separately.
%In this way the volume contains again the function $B(\gamma)$ defined in Eq.~\eqref{eq:definition_B_gamma}, since the $\varepsilon \rightarrow 0$ limit is regular and can be taken immediately. 
%The non-trivial reminders of this manipulation are evaluated using the identity
%\beq
%\int_1^{\frac{1}{\gamma \varepsilon^2}} d \tau \, \le \sqrt{\frac{\tau}{\gamma}} + \frac{1}{2 \gamma^{3/2} \sqrt{\tau}}  \ri =
%\frac{1}{\gamma^2} \le \frac{2}{3 \varepsilon^3} + \frac{1}{\varepsilon} - \sqrt{\gamma} - \frac{2}{3} \gamma^{3/2} \ri  \, ,
%\eeq
%which gives
\beq
\mathcal{V} = \frac{L^4 V_2}{2} \le \frac{1}{\delta^2} - \frac{1}{z_{\rm IR}^2}  \ri \le  \frac{2}{3 \varepsilon^3} + \frac{1}{\varepsilon}
+ \gamma^2 B(\gamma)
 - \sqrt{\gamma} - \frac{2}{3} \gamma^{3/2} \ri + \mathcal{O}(\varepsilon) \, . 
 \label{eq:extremal_volume_twocutoff}
\eeq 

\subsubsection*{Subtraction of vacuum $\mathrm{AdS}_5$}

With this other regularization, the cutoff along the $w$ coordinate is determined as
\beq
h(w^*) = \frac{1}{1-(w^*)^2} = \frac{1}{\varepsilon^2}  \quad
\Rightarrow \quad
w^* = \sqrt{1-\varepsilon^2} \, .
\eeq
Thus, the extremal volume for the undeformed case is
\beq
\mathcal{V}_{\mathrm{AdS}_5}  = 2 L^4 V_2 \int_{\delta}^{\infty} \frac{dz}{z^3} 
\int_0^{\sqrt{1-\varepsilon^2}} \frac{dw}{(1-w^2)^{5/2}} 
 = \frac{L^4 V_2}{\delta^2} \le \frac{1}{3 \varepsilon^3} + \frac{1}{2 \varepsilon} \ri + \mathcal{O} (\varepsilon) \, .
\label{eq:extremal_volume_emptyAdS_double_cutoff}
\eeq

\subsubsection*{Total result}

After subtracting the vacuum solution from the extremal volume \eqref{eq:extremal_volume_twocutoff} in the presence of the defect, we get the complexity of formation
\beq
\Delta \mathcal{C}_{\mathrm{AdS}_5} = \frac{V_2 L^3}{2 G}  \le \gamma^2 B(\gamma) - \sqrt{\gamma} - \frac{2}{3} \gamma^{3/2} \ri   \le  \frac{1}{\delta^2} - \frac{1}{z_{\rm IR}^2} \ri
\, .
\label{eq:complexity_formation_2cutoff}
\eeq
Comparing this result with Eq.~\eqref{eq:total_volume_JanusAdS5}, we notice that the coefficient of the divergent part is different, while the finite terms match.
This suggests that the universal information encoded by the complexity of formation is associated to the finite part, since a change of the energy scale does not affect it.
We should also  emphasize that the finite part is inversely proportional to the IR regulator, and then in the smooth limit $z_{\rm IR} \rightarrow \infty$ the corresponding expression vanishes. Therefore, no universal information is encoded in the complexity of formation for the entire Janus AdS$_5$ geometry, except that divergences scale as $\delta^{-2}.$

\section{Volume subregion for the Janus $ \mathrm{AdS}_5 $ geometry}
\label{sect-subregion_CV_AdS5} 

In this Section, we move to the case of subregion complexity to analyze if additional divergences (e.g., logarithms in the UV cutoff) arise in the computation of the extremal volume. If further types of divergences occur, we can get additional insights into the structure of universal terms in the complexity of formation.

In the CV case, subregion complexity is computed as
the volume of a maximal codimension-one bulk surface enclosed by the boundary subregion and the corresponding Hubeny-Rangamani-Takayanagi surface \cite{Alishahiha:2015rta}. 
In the CA case, it is computed as the gravitational action  in the intersection region between the WDW patch and the entanglement wedge built from the boundary subregion \cite{Carmi:2016wjl}. 
Subregion complexity has been investigated for asymptotically $\mathrm{AdS}_3$ spacetime, where it was found that the volume depends only on topological quantities but not on the temperature of the black hole \cite{Abt:2017pmf}.
Contrarily, the action for a generic segment on the boundary of the BTZ black hole is not topological, but is directly related to the corresponding entanglement entropy. 
This structure is spoiled when a larger number of intervals on the boundary is considered \cite{Auzzi:2019vyh}.
From the field theory point of view, subregion complexity is believed to be dual either to fidelity, complexity of purification or basis complexity \cite{Alishahiha:2015rta, MIyaji:2015mia, Agon:2018zso}.

\subsection{Ball-shaped subregion on the boundary}

\begin{figure}
	\begin{center}
		\begin{tikzpicture}[scale=1.2]
		
		\begin{axis}[axis equal,
		width=15cm,
		height=15cm,
		axis lines = center,
		xlabel = {$x_\perp$},
		ylabel = {$x_\parallel$},
		zlabel = {$z$},
		ticks=none,
		enlargelimits=0.3,
		z buffer=sort,
		view/h=45,
		scale uniformly strategy=units only]
		% this example burns colors if opacity 
		% is active in the document.
		
		\draw [black!30!white,fill] (-1.5,-1.5,0) -- (-1.5,1.5,0) -- (0,1.5,0) -- (0,-1.5,0) -- (-1.5,-1.5,0);
		\draw [black!10!white,fill] (1.5,-1.5,0) -- (1.5,1.5,0) -- (0,1.5,0) -- (0,-1.5,0) -- (-1.5,-1.5,0);
		\draw[black] (-1.6,0,0) -- (0,0,0);
		\draw[blue!80!green,very thick] (0,-1.5,0) -- (0,1.5,0);
		\addplot3 [patch,
		patch type=bilinear,
		mesh/color input=explicit mathparse,
		variable = \u,
		variable y = \v,
		domain = 0:360,
		y domain = 0:90,
		point meta={symbolic={0.5+0.5*y, % R 
				0.5+0.5*x, % G 
				0.5+0.5*z% B
		} },
		] ({cos(u)*sin(v)}, {sin(u)*sin(v)}, {cos(v)});
		\draw (1,0,0) -- (1.5,0,0)  (0,0,1)   -- (0,0,1.5);
		
		\end{axis}
		\end{tikzpicture}
		\caption{Time slice of the Janus $\mathrm{AdS}_5$ spacetime with a ball-shaped subregion centered on the interface. The Ryu-Takayanagi surface is represented by the spherical dome while the blue line represents the interface located at $x_\perp=0$.}
		\label{fig-ball_shaped_region_Janus_AdS5}
	\end{center}
\end{figure}
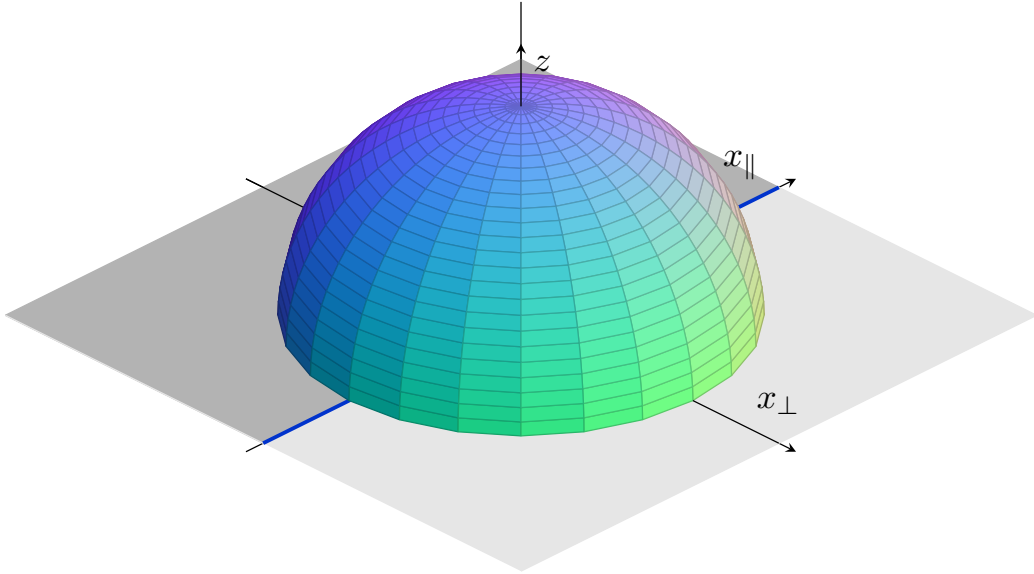

Following ideas similar to the ones used to investigate entanglement entropy in \cite{Estes:2014hka, Gutperle:2016gfe}, a particularly convenient scenario in which to study the subregion complexity, corresponds to a ball-shaped region of radius $R$ centered on the interface, see Fig.~\ref{fig-ball_shaped_region_Janus_AdS5}.
It is possible to show that the RT surface corresponding to such subsystem reads
\beq
\label{eq:geodesic_RT}
z^2 + r^2 = R^2 \, ,
\eeq
where this equation is written in terms of the metric \eqref{metric5} with the two-dimensional subspace parametrized with polar coordinates
\beq
d \vec{x}^2 = dr^2 + r^2 d \theta^2 \, .
\eeq
Remarkably, the space-like RT surface in Eq.~\eqref{eq:geodesic_RT} is the same found in empty $\mathrm{AdS}_5$ space. In fact, it can be shown that it also corresponds to a global minimum of the area functional in the Janus $\mathrm{AdS}_5$ geometry.

The RT surface delimits the region where the extremal codimension-one Cauchy slice, that computes the extremal volume, extends. Thus, the integration domain of the volume will be modified and will not reach $z=z_{\rm IR},$ as it happens instead in the full Janus $\mathrm{AdS}_5$ case.
Now, we need to evaluate
\beq
\mathcal{V}_{\rm sub} (R,\gamma) = \frac{2}{\sqrt{\gamma}} \int_0^{2 \pi} d \theta \int_{z_{\rm min}}^{z_{\rm max}} \frac{dz}{z^3}
\int_0^{\sqrt{R^2-z^2}} dr \, r \int_0^{w_{+}}  dw \, h^{\frac{5}{2}} (w) \, ,
\eeq
where we put a factor of 2 for symmetry reasons in the $w$ integration.
The integral along the angular direction is trivial, while the one along the radial direction of the polar coordinates gives an additional factor that will modify the last integration along $z.$
It is straightforward to find
\beq
\mathcal{V}_{\rm sub} (R,\gamma) = \frac{2 \pi}{\sqrt{\gamma}}
\int_{z_{\rm min}}^R dz \, \frac{R^2 -z^2}{z^3} 
\int_0^{w_{+}}  dw \, h^{\frac{5}{2}} (w) \, .
\label{eq:starting_point_subregion_volume}
\eeq
Notice that the maximum value that can be reached by $z$ is $R$. Otherwise, the square root defining the maximum of the radial coordinate $r$ would be imaginary.
The minimum of $z$ is determined in the same way as for the total volume, according to Eq.~\eqref{z_min_AdS_5} for the single cutoff prescription, whereas for the double cutoff regularization corresponds to $z_{\rm min}= \delta$ .

\subsection{Extremal volume: single cutoff procedure}
We start by computing the extremal volume for the ball-spahed region in the Janus $\mathrm{AdS}_5$ geometry using the single cutoff method. Once more it is useful to change variables in
\beq
 \tau=\frac{h(w)}{\gamma}, \qquad \zeta= \frac{z^2}{\gamma \delta^2}\,,
\eeq
so that the integral in Eq.~\eqref{eq:starting_point_subregion_volume} becomes
%\beq
%\mathcal{V}=\pi \gamma^2\int_{\sqrt{\gamma}\delta}^R dz \, \frac{R^2 -z^2}{z^3}\int_{1}^{\frac{z^2}{\gamma \delta^2}}\tau^{5/2}f(\tau)\;,
%\eeq
%where $f(\tau)$ is defined in Eq.~\eqref{def_func_f}. Performing a further change of variables $\zeta= z^2/(\gamma \delta^2)$, we are left with
\beq
\mathcal{V}_{\rm sub} (R,\gamma) =\pi L^4 \gamma^2 \int_{1}^{\frac{R^2}{\gamma \delta^2}} d\zeta \, \frac{R^2 -\gamma \delta^2 \zeta}{2\gamma\delta^2\zeta^2}\int_{1}^{\zeta}\tau^{5/2}f(\tau)\;.
\eeq
As explained in Eq.~\eqref{swapping} for the total Janus $\mathrm{AdS}_5$ space, we can swap the integrals in $\zeta$ and $\tau$ being mindful to perform the suitable changes in the extremes of integration, getting
\beq
\mathcal{V}_{\rm sub} (R,\gamma) =\pi L^4 \gamma^2\int_{1}^{\frac{R^2}{\gamma \delta^2}} \tau^{5/2}f(\tau)\int_{\tau}^{\frac{R^2}{\gamma \delta^2}}d\zeta \, \frac{R^2 -\gamma \delta^2 \zeta}{2\gamma\delta^2\zeta^2}\;.
\eeq  
Now, it is possible to evaluate first the integral in the $\zeta$ variable as
\beq
\int_{\tau}^{\frac{R^2}{\gamma \delta^2}}d\zeta \, \frac{R^2 -\gamma \delta^2 \zeta}{2\gamma\delta^2\zeta^2}=\frac{1}{2}\left(2\log\left(\frac{\sqrt{\gamma}\delta}{R} \right) -1+ \frac{R^2}{\gamma \delta^2 \tau}+\log \tau \right) \;.
\eeq  
Concerning the integration over $\tau$, we have to compute three different kinds of integrals. The first one can be evaluated following the same steps as in Section \ref{subsec_extremal_volume_AdS_5}
 \beq
\int_{1}^{\frac{R^2}{\gamma \delta^2}} \tau^{5/2}f(\tau)= B(\gamma)+\int_{1}^{\frac{R^2}{\gamma \delta^2}} \le \sqrt{\frac{\tau}{\gamma}} + \frac{1}{2 \gamma^{3/2} \sqrt{\tau}} \ri = B(\gamma)\, +\, \frac{2R^3}{3 \gamma^2\delta^3}+\frac {R}{2\gamma \delta^2} -\frac{1}{\gamma^{3/2}}-\frac{2}{3\sqrt{\gamma}}\;,
 \eeq
 where $B(\gamma)$ is defined in Eq.~\eqref{eq:definition_B_gamma} and we have taken the $\delta \rightarrow 0 $ limit. 
 
The second integral in $\tau$ is given by
\beq
\int_{1}^{\frac{R^2}{\gamma \delta^2}} \tau^{3/2}f(\tau)=A(\gamma) + \int_{1}^{\frac{R^2}{\gamma \delta^2}} \le \frac{1}{\sqrt{\gamma \tau}} + \frac{1}{2 \gamma^{3/2} \tau^{3/2}} \ri = A(\gamma)+\frac{2R}{\gamma \delta}+\frac{1}{\gamma^{3/2}}-\frac{2}{\sqrt{\gamma}}-\frac{\delta}{\gamma R}\;.
\eeq
Here, $A(\gamma)$ is defined as in Eq.~\eqref{eq:definition_A_gamma}. Since the result of the integral has to be multiplied by a factor of $R^2/(\gamma \delta^2)$ coming from the $\zeta$ integration, it is crucial to expand the above expression up to order $\delta$ to keep track of all the possible divergences.

The last type of integral in $\tau$ is
\beq
\begin{aligned}
\label{third_integral}
\int_{1}^{\frac{R^2}{\gamma \delta^2}} \tau^{5/2}\log \tau f(\tau) &= C(\gamma)+\int_{1}^{\frac{R^2}{\gamma \delta^2}}\le\frac{\log \tau \sqrt{\tau}}{\sqrt{\gamma}}+\frac{\log \tau }{2\gamma^{3/2}\sqrt{\tau}} \ri \\ &=C(\gamma) -\log \left(\frac{\sqrt{\gamma} \delta}{R} \right)\left( \frac{4R^3}{3\gamma^2\delta^3}+\frac{2R}{\gamma^2\delta}\right) -\frac{4R^3}{9\gamma^2\delta^3}-\frac{2R}{\gamma^2 \delta}+\frac{2}{\gamma^{3/2}}+\frac{4}{9\sqrt{\gamma}}\;,
\end{aligned}
\eeq
where we define
%The integrand can be rewritten as
%\beq
%\tau^{5/2}\log \tau f(\tau) = \tau^{5/2}\log \tau\left[\le  f(\tau) - \frac{1}{\sqrt{\gamma} \tau^2} - \frac{1}{2 \gamma^{3/2} \tau^3} \ri  + \frac{1}{\sqrt{\gamma} \tau^2} + \frac{1}{2 \gamma^{3/2} \tau^3}  \right] 
%\eeq
%in such a way that the term in the round brackets gives a finite integral. Defining
\beq
\label{eq:definition_C_gamma}
C(\gamma)= \int_{1}^{\infty} d\tau \, \tau^{5/2} \log \tau
\le  f(\tau) - \frac{1}{\sqrt{\gamma} \tau^2} - \frac{1}{2 \gamma^{3/2} \tau^3} \ri \, .
\eeq
This function can be evaluated numerically and it is analytic in all the range $3/4 \leq \gamma leq 1.$
Nevertheless, we will keep it implicit in the following manipulations.
%the integral presented in Eq.~\eqref{third_integral} becomes
%\beq
%C(\gamma)+\int_{1}^{\frac{R^2}{\gamma \delta^2}}\le\frac{\log \tau \sqrt{\tau}}{\sqrt{\gamma}}+\frac{\log \tau }{2\gamma^{3/2}\sqrt{\tau}} \ri 
%\eeq
%and finally gives
%\beq
%C(\gamma) -\log \left(\frac{\sqrt{\gamma} \delta}{R} \right)\left( \frac{4R^3}{3\gamma^2\delta^3}+\frac{2R}{\gamma^2\delta}\right) -\frac{4R^3}{9\gamma^2\delta^3}-\frac{2R}{\gamma^2 \delta}+\frac{2}{\gamma^{3/2}}+\frac{4}{9\sqrt{\gamma}}\;.
%\eeq
%The behavior of this function is shown in Fig.~\ref{fig-C_function}.
%\begin{figure}[ht]
%    \centering
%    \def\svgwidth{\columnwidth}
%    \scalebox{0.68}{\input{C_di_gamma.pdf_tex}}
%  	\caption{Numerical plot of the function $C(\gamma)$ defined in Eq.~\eqref{eq:definition_C_gamma}. Its value in the $\gamma \rightarrow 1$  limit is $C(1)\simeq1.74$.}
%	\label{fig-C_function}
%\end{figure} 

Combining together all the terms, we get the extremal subregion volume 
\beq
\label{subregion_1_cutoff}
\begin{aligned}
\mathcal{V}_{\rm sub} & (R,\gamma)  =\frac{4 \pi L^4}{9}\frac{R^3}{\delta^3}+\pi L^4 \frac{R^2}{\delta^2}\left(\frac{\gamma A(\gamma)}{2}\!-\!\sqrt{\gamma}+\frac{1}{2\sqrt{\gamma}} \right)\! -2\pi L^4 \frac{R}{\delta} \\
& +\pi L^4 \log \left(\frac{\sqrt{\gamma} \delta}{R} \right) \!\left(\gamma^2 B(\gamma) \!-\!\sqrt{\gamma} -\frac{2}{3}\gamma^{3/2}\right)
+\frac{\pi L^4}{2}\left( \gamma^2C(\gamma)-\gamma^2B(\gamma)+3\sqrt{\gamma}+\frac{10}{9}\gamma^{3/2}\right) \;.
\end{aligned}
\eeq

\subsubsection*{Subtraction of the vacuum AdS solution}
Now, we can perform the subtraction of the AdS vacuum solution, which corresponds to set $\gamma=w_0=1.$
The volume reads
\beq
\begin{aligned}
\mathcal{V}_{\rm sub} (R,\gamma=1)& = 2 \pi L^4 \int_{\delta}^R dz \, \frac{R^2 - z^2}{z^3} \int_0^{\sqrt{1-\frac{\delta^2}{z^2}}} \frac{dw}{(1-w^2)^{\frac{5}{2}}} = \\
& = \frac{4 \pi L^4}{9} \frac{R^3}{\delta^3} - 2 \pi L^4 \frac{R}{\delta} + \frac{2 \pi^2 L^4}{3} + \mathcal{O} (\delta) \, .
\end{aligned}
\eeq
\subsubsection*{Total result}
Subtracting the undeformed $\mathrm{AdS}_5$ solution to the extremal volume obtained in Eq.~\eqref{subregion_1_cutoff}, we find that the subregion complexity of formation is given by
\beq
\begin{aligned}
\Delta \mathcal{C}_{\rm{sub}} (R,\gamma) =&\frac{\pi L^3}{G}\frac{R^2}{\delta^2}\left(\frac{\gamma A(\gamma)}{2}\!-\!\sqrt{\gamma}+\frac{1}{2\sqrt{\gamma}} \right)\!+\frac{\pi L^3}{G} \log \left(\frac{\sqrt{\gamma} \delta}{R} \right) \!\left(\gamma^2 B(\gamma) \!-\!\sqrt{\gamma} -\frac{2}{3}\gamma^{3/2}\right)+ \\&+\frac{\pi L^3}{2G}\left( \gamma^2C(\gamma)-\gamma^2B(\gamma)+3\sqrt{\gamma}+\frac{10}{9}\gamma^{3/2}-\frac{4\pi}{3}\right) \;.
\end{aligned}
\label{complexity_pf_fomration_subregion_1_cutoff}
\eeq
A comment on the limit $R \rightarrow \infty$ is in order. This limit corresponds to the case where the subregion covers the entire boundary, and should therefore map to the result for the total volume computed in Eq.~\eqref{eq:total_volume_JanusAdS5}.
In order to verify that this is the case, we notice that in polar coordinates the two-dimensional volume along the spatial directions $\vec{x} = (r, \theta)$ is
\beq
V_2 = \pi R^2 \, ,
\eeq
which becomes infinite in the limit $R \rightarrow \infty. $
For this reason, when comparing the two quantities we should check that
\beq
\frac{\Delta \mathcal{V}}{V_2} = \lim_{R \rightarrow \infty} \frac{\Delta \mathcal{V}_{\rm sub}}{\pi R^2} \, .
\label{eq:limit_subregion_bigR}
\eeq
The limit in the RHS of Eq.~\eqref{eq:limit_subregion_bigR} suppresses logarithmic and finite terms, and allows only to compare the divergent parts which are proportional to the volume of the subregion.
%\footnote{Notice that the same observations held when comparing the complexity=action for the BTZ black hole when the subregion on the boundary is taken was sent to infinity \cite{Auzzi:2019vyh}.}.
Employing Eq.~\eqref{eq:limit_subregion_bigR}, we immediatly recognize that the terms proportional to $\delta^{-2}$ in Eq.~\eqref{eq:total_volume_JanusAdS5} and \eqref{complexity_pf_fomration_subregion_1_cutoff} exactly match.
In conclusion, the difference between the total and the subregion case for the complexity=volume conjecture involving the Janus deformation of $\mathrm{AdS}_5$ spacetime, amounts to the presence of an additional finite term and a logarithmic divergence.

\subsection{Extremal volume: double cutoff procedure}

We evaluate the volume in Eq.~\eqref{eq:starting_point_subregion_volume} using the double cutoff regularization scheme, which consists in setting $z_{\rm min} = \delta.$ The integral along the $z$ variable reads
\beq
\int_{\delta}^R dz \, \frac{R^2-z^2}{z^3} = \frac{R^2}{2 \delta^2} + \log \le \frac{\delta}{R} \ri - \frac{1}{2} \, .
\eeq
After the change of variables $\tau = \gamma^{-1} h(w),$ the volume becomes
\beq
\begin{aligned}
\mathcal{V}_{\rm sub}(R,\gamma) & = \pi L^4 \gamma^2 \le \frac{R^2}{2 \delta^2} + \log \le \frac{\delta}{R} \ri - \frac{1}{2}  \ri 
\int_1^{\frac{1}{\gamma \varepsilon^2}} d\tau \, \tau^{5/2} f(\tau) =\\ &=  \pi L^4 \le \frac{R^2}{2 \delta^2} + \log \le \frac{\delta}{R} \ri - \frac{1}{2} \ri
\le \gamma^2 B(\gamma) + \frac{2}{3 \varepsilon^3} + \frac{1}{\varepsilon} - \sqrt{\gamma} - \frac{2}{3} \gamma^{3/2}  \ri + \mathcal{O} (\varepsilon) \,.
\end{aligned}
\label{eq:result_volume_subregion_2cutoff}
\eeq
%The integral over $\t$ is equivalent to the one considered in Section \ref{sect-Extramal_volume_2cutoff}, and then it simply gives
%\beq
%\mathcal{V}_{\rm sub} = \pi \le \frac{R^2}{2 \delta^2} + \log \le \frac{\delta}{R} \ri - \frac{1}{2} \ri
%\le \gamma^2 B(\gamma) + \frac{2}{3 \varepsilon^3} + \frac{1}{\varepsilon} - \sqrt{\gamma} - \frac{2}{3} \gamma^{3/2}  \ri + \mathcal{O} (\varepsilon) \, .
%\label{eq:result_volume_subregion_2cutoff}
%\eeq
\subsubsection*{Subtraction of the vacuum AdS solution}
The corresponding volume in the empty AdS geometry is easily obtained by considering 
\beq
\begin{aligned}
\mathcal{V}_{\mathrm{sub}} (R,0) &= 2 \pi L^4 \int_{\delta}^R dz \, \frac{R^2-z^2}{z^3} 
\int_0^{\sqrt{1-\varepsilon^2}} \!\!\!\frac{dw}{(1-w^2)^{5/2}} =\\&=
2 \pi L^4 \le \frac{R^2}{2 \delta^2} + \log \le \frac{\delta}{R} \ri - \frac{1}{2} \ri 
\le \frac{1}{3 \varepsilon^3} + \frac{1}{2 \varepsilon}  \ri + \mathcal{O}(\varepsilon) \, .
\end{aligned}
\eeq
\subsubsection*{Total result}
After subtracting the vacuum solution from Eq.~\eqref{eq:result_volume_subregion_2cutoff}, we get the complexity of formation in the double cutoff regularization scheme
\beq
\Delta \mathcal{C}_{\rm sub} (R,\gamma)= \frac{\pi L^3}{G} \le \frac{R^2}{2 \delta^2} + \log \le \frac{\delta}{R} \ri - \frac{1}{2} \ri 
\le \gamma^2 B(\gamma) - \sqrt{\gamma} - \frac{2}{3} \gamma^{3/2}  \ri + \mathcal{O}(\varepsilon) \, .
\eeq
It can be easily checked that in the $R \rightarrow \infty$ limit, one finds 
\beq
\lim_{R \rightarrow \infty} \frac{V_2 L^4}{\pi R^2} \, \Delta \mathcal{C}_{\rm sub}(R,\gamma)  = 
\frac{V_2}{2 G \,\delta^2} \le \gamma^2 B(\gamma) - \sqrt{\gamma} - \frac{2}{3} \gamma^{3/2}  \ri \, ,
\eeq
which matches the $\delta^{-2}$ divergent part in  Eq.~\eqref{eq:complexity_formation_2cutoff}.

\section{Conclusions}
\label{sect-conclusions}

%\begin{figure}[ht]
%	\centering
%	\includegraphics[scale=0.47]{Universal_coefficient.pdf}
%	\caption{Numerical plot of the coefficient of the universal logarithmic divergence in the subregion volume.}
%	\label{fig-universal_coefficient}
%\end{figure}

\begin{figure}[ht]
    \centering
    \def\svgwidth{\columnwidth}
    \scalebox{0.7}{%% Creator: Inkscape 1.0.2-2 (e86c870879, 2021-01-15), www.inkscape.org
%% PDF/EPS/PS + LaTeX output extension by Johan Engelen, 2010
%% Accompanies image file 'FandG.pdf' (pdf, eps, ps)
%%
%% To include the image in your LaTeX document, write
%%   \input{<filename>.pdf_tex}
%%  instead of
%%   \includegraphics{<filename>.pdf}
%% To scale the image, write
%%   \def\svgwidth{<desired width>}
%%   \input{<filename>.pdf_tex}
%%  instead of
%%   \includegraphics[width=<desired width>]{<filename>.pdf}
%%
%% Images with a different path to the parent latex file can
%% be accessed with the `import' package (which may need to be
%% installed) using
%%   \usepackage{import}
%% in the preamble, and then including the image with
%%   \import{<path to file>}{<filename>.pdf_tex}
%% Alternatively, one can specify
%%   \graphicspath{{<path to file>/}}
%% 
%% For more information, please see info/svg-inkscape on CTAN:
%%   http://tug.ctan.org/tex-archive/info/svg-inkscape
%%
\begingroup%
  \makeatletter%
  \providecommand\color[2][]{%
    \errmessage{(Inkscape) Color is used for the text in Inkscape, but the package 'color.sty' is not loaded}%
    \renewcommand\color[2][]{}%
  }%
  \providecommand\transparent[1]{%
    \errmessage{(Inkscape) Transparency is used (non-zero) for the text in Inkscape, but the package 'transparent.sty' is not loaded}%
    \renewcommand\transparent[1]{}%
  }%
  \providecommand\rotatebox[2]{#2}%
  \newcommand*\fsize{\dimexpr\f@size pt\relax}%
  \newcommand*\lineheight[1]{\fontsize{\fsize}{#1\fsize}\selectfont}%
  \ifx\svgwidth\undefined%
    \setlength{\unitlength}{749bp}%
    \ifx\svgscale\undefined%
      \relax%
    \else%
      \setlength{\unitlength}{\unitlength * \real{\svgscale}}%
    \fi%
  \else%
    \setlength{\unitlength}{\svgwidth}%
  \fi%
  \global\let\svgwidth\undefined%
  \global\let\svgscale\undefined%
  \makeatother%
  \begin{picture}(1,0.66755674)%
    \lineheight{1}%
    \setlength\tabcolsep{0pt}%
    \put(0,0){\includegraphics[width=\unitlength,page=1]{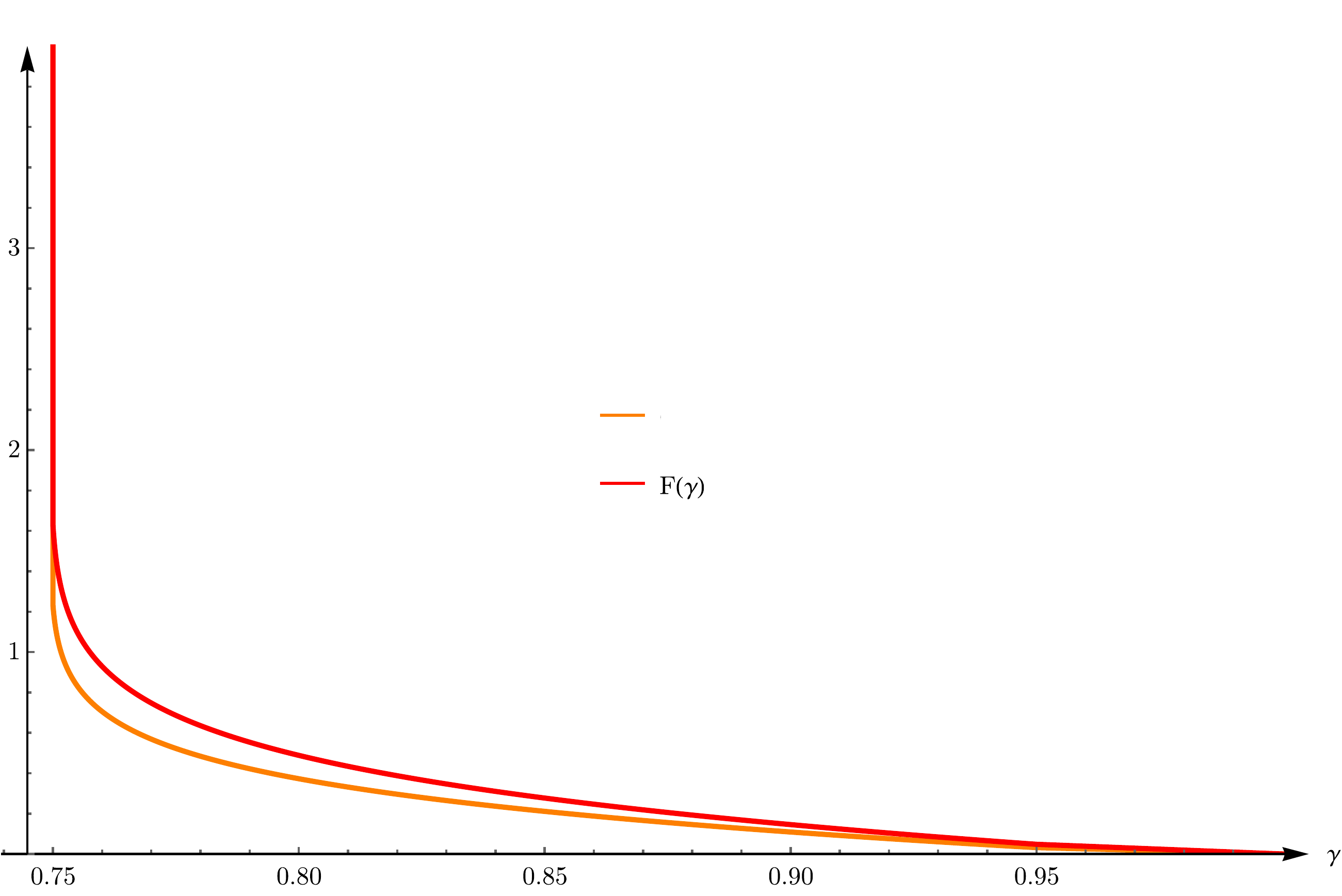}}%
    \put(0.48738924,0.3479386){\makebox(0,0)[lt]{\lineheight{1.25}\smash{\begin{tabular}[t]{l}$\mathcal{G}(\gamma)$\end{tabular}}}}%
    \put(0,0){\includegraphics[width=\unitlength,page=2]{FandG.pdf}}%
    \put(0.48817765,0.29848183){\makebox(0,0)[lt]{\lineheight{1.25}\smash{\begin{tabular}[t]{l}$\mathcal{F}(\gamma)$\end{tabular}}}}%
    \put(0,0){\includegraphics[width=\unitlength,page=3]{FandG.pdf}}%
    \put(0.97757771,0.02372461){\makebox(0,0)[lt]{\lineheight{1.25}\smash{\begin{tabular}[t]{l}$\gamma$\end{tabular}}}}%
  \end{picture}%
\endgroup%
}
  	\caption{Numerical plot of the functions $\mathcal{F}(\gamma)$ and $\mathcal{G}(\gamma)$ defined in Eq.~\eqref{eq:F_and_G}. Both functions vanish in the limit $\gamma\rightarrow 1$.}
	\label{fig-FandG_function}
\end{figure}

We computed the holographic complexity of formation for the Janus deformation of $\mathrm{AdS}_5$ spacetime with respect to vacuum space both for the entire boundary and for the case of a symmetric ball-shaped subregion located symmetrically around the interface.
We employed two different prescriptions to regularize the extremal volume, i.e., the single and double cutoff methods.
The coefficients of the UV divergences are collected in Table \ref{tab:results}, expressed in terms of the following functions 
\beq
\mathcal{F} (\gamma) \equiv  \frac{\gamma A(\gamma)}{2} - \sqrt{\gamma}+ \frac{1}{2\sqrt{\gamma}}  \, ,  \qquad
\mathcal{G} (\gamma) \equiv \frac{\gamma^2 B(\gamma)}{2} - \frac{\sqrt{\gamma}}{2} - \frac{1}{3} \gamma^{3/2} \, .
\label{eq:F_and_G}
\eeq
These functions are depicted in fig. \ref{fig-FandG_function}.

\begin{table}[ht]   
\begin{center}    
\begin{tabular}  {|c|c|c|} \hline Complexity of formation & Single cutoff & Double cutoff  \\ \hline
\rule{0pt}{4.9ex}  Entire boundary    & $ \displaystyle \frac{L^3}{G} \frac{V_2}{\delta^2} \, \mathcal{F}(\gamma) $  & $ \displaystyle \frac{L^3}{G} \frac{V_2}{\delta^2} \,  \mathcal{G}(\gamma)  $ \\ 
\rule{0pt}{4.9ex} Subregion: $\delta^{-2}$ term &  $ \displaystyle \frac{L^3}{G}\frac{\pi R^2}{\delta^2} \, \mathcal{F}(\gamma) \!$  &  $ \displaystyle  \frac{L^3}{G}  \frac{\pi R^2}{\delta^2} \, \mathcal{G}(\gamma)    $   \\
\rule{0pt}{4.9ex} Subregion: $\log \delta $ term  &  $ \displaystyle \frac{2 \pi L^3}{G} \log \left(\frac{\sqrt{\gamma} \delta}{R} \right) \! \mathcal{G}(\gamma)  $ & $ \displaystyle  \frac{2 \pi L^3}{G} \log \le \frac{\delta}{R} \ri  \mathcal{G}(\gamma) $  \\ 
\hline
\end{tabular}   
\caption{Coefficients of the divergences entering the complexity of formation for the non-SUSY Janus $\mathrm{AdS}_5$ geometry.
\label{tab:results}}
\end{center}
\end{table}

First of all, we notice that the structure of divergences differs between the entire boundary case and the subregion setting.
The former is characterized by a power-law divergence $\delta^{-2}$, which is consistent with the result computed in \cite{Sato:2019kik} for a BCFT.
The subregion complexity, on the other hand, has a richer structure, where an additional logarithmic divergence and a non-vanishing finite terms appears.
A similar difference is also present in the complexity=action computation involving the (2+1)-dimensional vacuum AdS or BTZ black hole solutions \cite{Auzzi:2019vyh}. 

Comparing the entries in Table \ref{tab:results}, the only result independent of the regularization scheme is the coefficient of the logarithmic divergences in the subregion case\footnote{Notice that the results only differ by the term $\frac{\pi L^3}{G} \log \gamma,$ which amounts to a finite part.}.
In addition, the finite term in the total volume case also matches between single and double cutoff prescriptions, see Eqs.~\eqref{eq:total_volume_JanusAdS5} and \eqref{eq:complexity_formation_2cutoff}, but it vanishes once we take the limit $z_{\rm IR} \rightarrow \infty$ for the IR regulator.  
This behavior suggests that similarly to the entanglement entropy computation, universal properties about complexity are encoded by logarithmic or finite terms since they are invariant under rescalings of the UV cutoff. 
When both terms are present, only the coefficient of the logarithm is universal. In fact, a transformation of the UV cutoff in the logarithm amounts to an additional finite part, which then becomes ambiguous.
Such remark is also consistent with the three-dimensional case considered in \cite{Auzzi:2021nrj}, where the complexity of formation was composed by a logarithmic divergence and a finite term in the Janus $\mathrm{AdS}_3$ and in the static Janus BTZ backgrounds.
In both cases, distinct regularizations differ by the finite part, but lead to the same coefficient of the logarithmic divergence.

There are some natural developments of this work that we aim to investigate.
A classification of the UV divergences for the complexity=action conjecture in the Janus $\mathrm{AdS}_5$ background would shed more light on the persistency of universality encoded by the logarithmic or finite terms in the volume case. 
Moreover, this will allow us to compare the UV divergences between the volume and the action cases in a higher dimensional defect geometry, as was accomplished in \cite{Sato:2019kik} for a BCFT.
Similarly to their approach, a computation on the field theory side would guide us towards a deeper understanding of the properties of complexity in the presence of an interface. 
One can perform this investigation using the path integral approach \cite{Caputa:2017yrh}, or by studying the geometry that arise in a dual CFT where some of its symmetries are broken, generalizing the method used in \cite{Chagnet:2021uvi} to the case of interfaces.
The computation for the extremal volume can also be pursued for the higher dimensional generalizations of the time-dependent Janus BTZ black hole proposed in \cite{Bak:2007qw}.

\section*{Acknowledgments}

We thank Roberto Auzzi and Giuseppe Nardelli for many valuable discussions.
The authors acknowledge support from the Independent Research Fund Denmark grant number DFF-6108-00340 “Towards a deeper understanding of black holes with non-relativistic holography” and from DFF-FNU through grant number DFF-4002-00037. 

\appendix
\section{Weierstrass $\wp$ function}
\label{app-weierstrass}

The Weierstrass $\wp$ is an elliptic function of order 2 defined by the series
\beq
\wp (z, \omega_1, \omega_2) = \frac{1}{z^2} + 
\sum_{(m, n) \ne (0,0)} \left[ \frac{1}{(z - 2 m \omega_1 -2 n \omega_2)^2} - \frac{1}{(2m \omega_1 + 2 n \omega_2)^2}   \right] \, ,
\eeq 
which is doubly periodic in the complex plane with half-periods $\omega_1, \omega_2 .$ 
It is a meromorphic and even function of $z$ with double poles at the lattice point defined by its periods.
One can alternatively define the elliptic $\wp$--function in terms of its invariants $g_2, g_3,$ which can be computed as Eisenstein series involving the half-periods $\omega_1, \omega_2 .$ 
However, in this case it is simpler to define the $\wp$--function as the solution to the differential equation
\beq
(\p_w \wp)^2 = 4 \wp^3 - g_2 \wp - g_3 \, .
\label{eq:differential_eq_WeierstrassP}
\eeq
We also define the Weierstrass $\zeta$ and $\sigma$--functions as
\beq
\wp(z) = - \zeta'(z) \, , \qquad
\zeta (z) = \frac{\sigma'(z)}{\sigma(z)} \, .
\label{eq:zeta_sigma_functions}
\eeq

\bibliography{at}
\bibliographystyle{at}

\end{document}